# A corner-cube-feedback Faraday laser with 8 kHz linewidth

Zhiyang Wang,[1] Zijie Liu,[1] Jianxiang Miao,[1] Hangbo Shi,[1] Xiaomin Qin,[1] Xiaolei Guan,[1] Jia Zhang,[1] Pengyuan Chang,[2] Tiantian Shi,[3,4*] and Jingbiao Chen[1,5]

[1]*State Key Laboratory of Advanced Optical Communication Systems and Networks, School of Electronics, Peking University, Beijing 100871, China*
[2]*Institute of Quantum Information and Technology, Nanjing University of Posts and Telecommunicatios, Nanjing 210003, China*
[3]*School of Integrated Circuits, Peking University, Beijing 100871, China*
[4]*National Key Laboratory of Advanced Micro and Nano Manufacture Technology, Beijing 100871, China*
[5]*Hefei National Laboratory, Hefei 230088, China*
*\*Corresponding author: tts@pku.edu.cn*

**Abstract:** A single-mode Cs atom 852 nm Faraday laser based on the corner cube feedback is demonstrated, and termed as corner-cube-feedback Faraday laser. Using the corner-cube retroreflector as external cavity feedback element in Faraday laser, mechanical robustness can be greatly improved due to the precise reflection of the incident light beam back to its original direction. This Faraday laser can achieve laser oscillation at a large angle, which between the incident light and the optical axis of corner cube, ranging from +3° to -3°. The most probable linewidth is 8 kHz measured by heterodyne beating with two identical lasers. Moreover, its output frequency remains close to the Cs atomic Doppler-broadened transition line, even though the diode current changes from 55 mA to 155 mA and the diode working temperature varies from 11.8°C to 37.2°C. The corner-cube-feedback Faraday laser with high mechanical robustness as well as narrow linewidth can be widely used in quantum precision measurement, such as atomic clocks, atomic gravimeters, and atomic magnetometers, etc.

## 1. Introduction

The external cavity diode lasers (ECDLs), whose frequency corresponds to the atomic transition lines, are greatly important in many fields of fundamental physical research, especially in atomic physics [1,2], atomic frequency standards [3], atomic magnetometers [4], laser cooling [5], and atomic gravimeters [6]. Traditionally, the frequency selective methods for ECDLs are gratings [7-9], Fabry-Pérot etalons [10,11] and interference filters [12,13]. The ECDLs based on those methods always have good tunability, and these methods are commonly used in commercial products. While the output wavelength of the ECDLs mentioned above cannot correspond to the atomic lines immediately when they are powered on, and the output wavelength is also easily affected by fluctuations in the temperature and current of the laser diode. The Faraday anomalous dispersion optical filter (FADOF), which was first reported in 1956 [14], can also be used to achieve frequency selection for the new generation of ECDLs. The ECDL, which uses FADOF as the frequency selectivity element and combine with the antireflection-coated laser diode as the gain medium, was first achieved [15] and has also globally recognized as "Faraday laser" [16-18]. These lasers have extremely good frequency performance and are robust to changes in the current and working temperature of laser diode. In the FADOF, the Faraday anomalous dispersion effect only occurs when the frequency of the incident light is near the resonant transition frequency of the atoms [19-22]. As a result, the FADOF can limit the laser frequency to the atomic Doppler-broadened line. Combining this with the use of antireflection-coated laser diode to overcome the disadvantage of the internal-cavity modes [23] leads to the remarkable phenomenon where Faraday lasers exhibit immunity to fluctuations in the laser diodes' working temperature and driving current [15,18]. Their output wavelength is always within the vicinity of the atomic transition lines as soon as they are powered on.

Due to its excellent performance, Faraday lasers are widely used in a lot of researches, such as single-mode laser [17], dual-mode laser [24], optical communication [25], high output power laser [26-28], high frequency stability laser[29,30]. In the above applications, the planar reflecting mirror is normally employed as cavity feedback element, which means that the light must be incident perpendicular to the surface of the planar reflecting mirror to establish laser oscillation, leading the Faraday laser is not robust enough against mechanical vibrations, then widen laser linewidth. Compared to the precise optical path alignment required for light reflected by a planar mirror, employing a cavity feedback element (such as a corner-cube retroreflector) that can always reflect the injected light back to its source—even in the presence of mechanical vibrations—ensures that laser oscillation can still be achieved [31]. As a result, this configuration endows the laser with strong adaptability to environmental changes.

In this paper, we demonstrate a Faraday laser based on a corner-cube retroreflector as the cavity feedback element, whose mechanical robustness and reliability have been improved, and the linewidth is also narrowed benefiting from this. Taking advantage of the corner-cube retroreflector, this Faraday laser can always perform normally until the angle between incident light and the axis of corner-cube retroreflector is beyond the ±3° range. The maximum frequency drift range at +3° is 458 MHz within 9 hours when the driving current and working temperature of laser diode are not altered. Because of the improvement of mechanical robustness, the most probable linewidth of Faraday laser is 8 kHz, which is narrower than all the Faraday lasers that have been reported. Maintaining the same advantage as all Faraday lasers, its output frequency aligns with the Cs atomic transition line across the laser diode's operating current range of 55 mA to 155 mA and working temperature range of 11.8°C to 37.2°C. Combining the Faraday laser's immunity to fluctuations of driving current and working temperature with the improved mechanical robustness and narrower linewidth brought by the corner-cube retroreflector, it is believed that this corner-cube Faraday laser can effectively enhance the system performance in quantum precision measurement applications.

## 2. Experimental Setup

The experimental setup of the Faraday laser is shown in Fig. 1(a). As shown in Fig. 1(a), the Faraday laser is mainly composed of an antireflection-coated laser diode (Toptica, EYP-RWE-0860-06010-1500-SOT02-0000) which is used as the gain medium, a FADOF which can achieve frequency selection in ECDLs, and a cavity mirror which can provide the feedback to establish laser oscillation. The lens is used to collimate the light from the laser diode, and the piezo transducer is used to achieve cavity tunability. To upgrade the mechanical robustness of the Faraday laser, a corner-cube retroreflector (Thorlabs, HHR1272-M03), which has beam deviations of less than 2 arcseconds, was used as the cavity mirror feedback element in this work. Figure 1(a) shows the Faraday laser without the rotation stage. We describe this configuration as the "original position", where the optical axis of the corner-cube retroreflector and the incident light overlap with each other. In addition, under this operation condition, parameters such as the output power with current, the output mode with current, the relative intensity noise (RIN), the phase noise, the linewidth and the frequency stability were measured to characterize the laser. We also fixed the corner-cube retroreflector on a rotation stage (Thorlabs, PR01/M) to establish laser oscillation, and tested parameters including wavelength range with current, wavelength range with time, as well as output mode change with current under different rotation angles. The cavity lengths are approximately 35 cm with the rotation stage and 23 cm without it, which are longer than those of other commercial ECDLs that with millimeter-level cavities. It should be noted that the vertex of the corner-cube retroreflector must be aligned as closely as possible with the axis of rotation stage, thus the rotation angle of the rotation stage can represent the rotation angle between the incident light and the optical axis of the corner-cube retroreflector. For comparison, the current-wavelength curve was also measured in the original position.

The output wavelength is measured by a wavelength meter (Bristol, 671A), whose accuracy and repeatability is ± 0.2 parts per million and ± 0.03 parts per million, respectively. The output power and the output mode are measured at the same time with wavelength. A power meter (Thorlabs, PM100D and S121C) was used to measure the output power, and a Fabry-Pérot interferometer (Thorlabs, SA210-8B) was used to scan the output mode for every driving current point we set. We used a spectrum analyzer (Keysight, N9010B) to estimate the beating linewidth of two Faraday lasers both based on 1000 G through heterodyne beating. The frequency stability of the laser during free-running operation was assessed by measuring the beat note signal with a frequency counter (Keysight, 53230A). The RIN and the phase noise were also measured during free-running operation by a phase noise analyzer (Rohde&Schwarz, FSWP), the baseband and RF interface were utilized, respectively. Figure 1(b) is the relevant Cs energy diagram of this paper, specifically, the $6S_{1/2}$ $F=4$ → $6P_{3/2}$ $F'$ are the transition lines we focused. Ideally, the output wavelength should be confined to a narrow frequency range close to those transition lines, depending on the bandwidth of the transmission spectrum of FADOF.

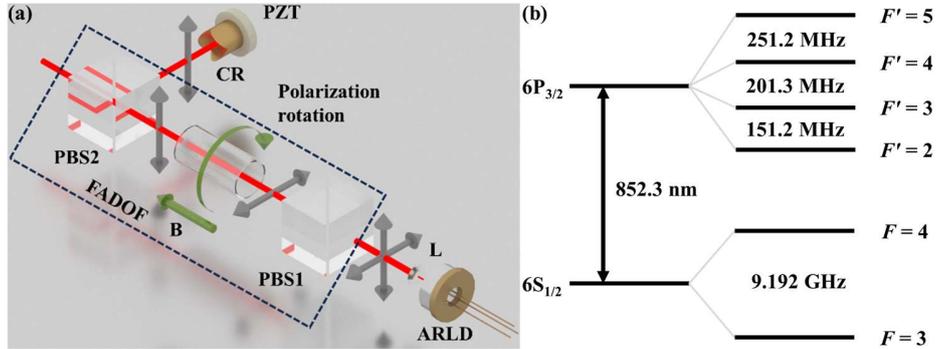

Fig. 1. (a) Experimental setup of Faraday laser without rotation stage. The cavity length is about 35 cm when the rotation stage is present and about 23 cm when it is absent. The double-arrows symbolize the direction of polarization of the light. ARLD: Antireflection-coated laser diode; L: Lens; PBS1 and PBS2: Polarizing beam splitter; B: Permanent magnet; CR: Corner-cube-retroreflector; PZT: Piezo transducer; (b) Relevant Cs energy level diagram.

## 3. Transmission spectrum of FADOF

The transmission spectrum profile of FADOF is very important for achieving single-mode Faraday laser with the wide variation of laser diodes' driving current. As shown in Fig. 1(a), a FADOF usually contains two polarizing beam splitters (PBSs). The alkali vapor cell is sandwiched between the two PBSs. Because the polarization directions of the two PBSs are perpendicular to each other, only the incident light whose rotation angle of polarization direction caused by Faraday effect is close to 90° can pass through the second PBS. The transmission spectrum profile is related to the length of the alkali vapor cell, the strength of the magnetic field, and the atomic density in the alkali vapor cell (which is determined by the temperature of the alkali vapor cell). Besides, the light intensity is also another decisive factor, and it is hard to maintain the transmission spectrum profile when the incident intensity changes from weak to strong. Fig. 2(a) is the saturated absorption spectroscopy (SAS) of Cs-$D_2$-line, and it can be used as the reference for the transmission spectrum profile, so we can know the output frequency of the Faraday laser when the FADOF is used as the frequency selective element. To optimize the transmission spectrum profile for consistently achieving single-mode laser output, we measured it under various working parameters. These included a magnetic field range from 300 G to 2000 G and a vapor-cell temperature from 40°C to 80°C, using a 30 mm-long vapor cell. There are several choices that have the potential to establish laser oscillation. The transmission spectrum profile of the parameters we chose from all the measured results are shown in Fig. 2(b), whose working parameters are 30 mm-long vapor cell, 1000 G magnetic field, and 56.1 °C vapor-cell temperature, respectively. The zero point in this figure refers to the $6S_{1/2}$ $F=4$ → $6P_{3/2}$ $F'=3$ transition (the output wavelength is around 852.3574 nm). We can see that those parameters we use can restrain the influence of the light intensity to the transmission spectrum profile, so the maximum transmission spectrum profile remains almost unchanged when the light intensity changes from weak to strong. Benefitting from this, the Faraday lasers' output wavelength can easily correspond to the Cs atomic Doppler-broadened transition line.

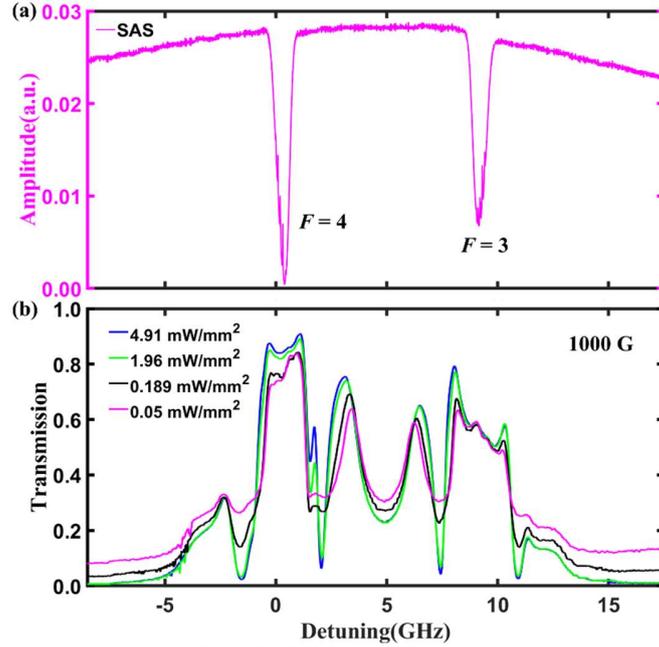

Fig. 2. Transmission spectrum of the FADOF. (a) The saturated absorption spectroscopy of $^{133}$Cs 852 nm. (b) Transmission spectrum of the FADOF with 30 mm-long vapor cell, 1000 G magnetic field, and 56.1 °C vapor-cell temperature.

## 4. Results and Discussion

### A. Results in original position

Figure 3(a) displays both the current-wavelength curve and the current-power curve for the original position. According to the datasheet of the ARLD, the recommended maximum operational driving current is 180 mA. Therefore, we set the driving current to range from 55 mA to 155 mA (a homemade current source), increasing in increments of approximately 2 mA. Mode hopping was observed at three points: around 81 mA, 103 mA, and 142 mA, which correspond to frequency shifts of approximately 408 MHz, 783 MHz, and 288 MHz, respectively. The linear fit slopes between the wavelength and the driving current, calculated among different mode hopping instances, are respectively 0.045 pm/mA, 0.052 pm/mA, 0.048 pm/mA, and 0.042 pm/mA, indicating a relatively consistent trend. We can also observe that, even when mode hopping occurs at certain points, the wavelength still remains close to the atomic transition lines ($6S_{1/2}$ $F=4$ → $6P_{3/2}$ $F'=5$, the corresponding wavelength is about 852.3563 nm). This stability significantly facilitates frequency locking via atomic or molecular methods and is urgently needed for many practical applications, especially in quantum-related areas. As for the output power, it naturally increases with the current within the same mode and undergoes related changes during mode hopping. Linear regression analysis indicates that the slopes, associated with different instances of mode hopping, are 0.473 mW/mA, 0.594 mW/mA, 0.804 mW/mA, and 0.879 mW/mA, respectively. Unlike the relatively consistent trend observed between output wavelength and driving current, these slopes exhibit a growing trend. We hypothesize that this is due to an increase in the transmittance of the FADOF as the intensity of the injected light increases, as demonstrated in Fig. 2(b). Consequently, the slope demonstrates a growing trend. Figure 3(b) shows the output wavelength versus the working temperature of the ARLD. The wavelength can always remain within the $6S_{1/2}$ $F$=4 → $6P_{3/2}$ $F$"=5 transition line when the temperature changes from 11.8 °C to 37.2 °C. Figure 3(c) depicts the output wavelength varying with the driving voltage on the PZT. Mode hopping occurs when the driving voltage exceeds the laser mode's stable range, which is approximately 7V. This corresponds to the biggest mode-hop-free range around 810 MHz. From Fig. 3(c), it is obvious that the output wavelength consistently remains close to the $6S_{1/2}$ F=4 → $6P_{3/2}$ F'=5 transition line, even when mode hopping occurs.

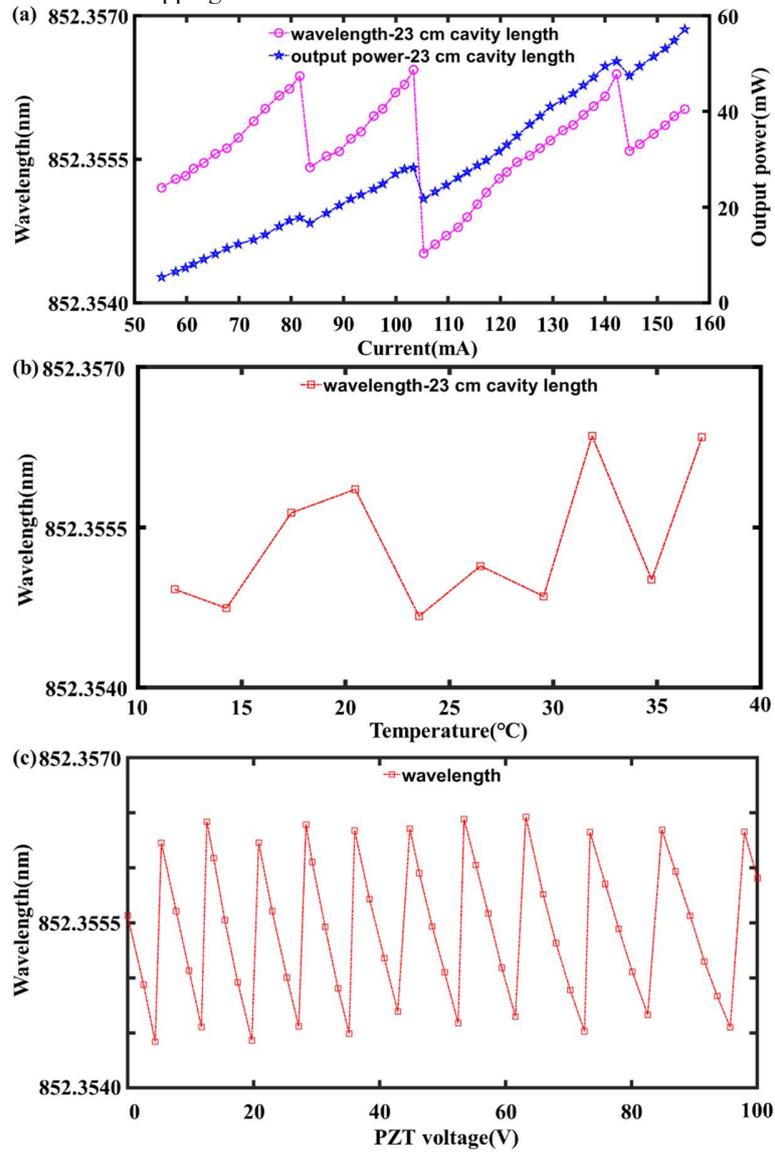

Fig. 3. The laser performance when the corner-cube retroreflector is at the original position. (a) The current-wavelength curve and the current-output power curve. (b) The temperature-wavelength curve, and we kept the driving current as a constant at 114.1 mA. (c) The output wavelength versus driving voltage on the PZT. The driving current and working temperature of the ARLD were set to 114.1 mA and 26.7°C, respectively.

**B. Results with rotation stage**

To demonstrate the excellent mechanical robustness of this corner-cube feedback Faraday laser, we installed the corner-cube retroreflector on a rotation stage with about 35 cm cavity length to establish laser oscillation. In order to determine the initial position of the rotation stage, the light was incident vertically to the vertex of the corner-cube retroreflector. This can ensure that the incident and reflected light overlap for at least two different spatial positions. Under this condition, the incident light aligns with the optical axis of the corner-cube retroreflector. We define this alignment as the 0° position. Subsequently, we rotated the rotation stage to various angles to observe the laser oscillation behavior. Since the vertex of the corner-cube retroreflector was aligned as precisely as possible with the axis of the rotation stage, the rotation angle can accurately reflect the angle between the incident light and the axis of the corner-cube retroreflector.

Figure 4(a) presents the current-wavelength curve for the laser operating at different rotation angles. Evidently, the laser oscillation is only disrupted when the angle between incident light and the optical axis of the corner-cube retroreflector exceeds the ±3° range. When the rotation angle is 0°, the oscillation current range is from 40.8 mA to 156.6 mA, and when the rotation angle is ±1°, ±2°, the oscillation current ranges are from 57.7 mA to 155.4 mA, 51.8 mA to 154.7 mA, 61.6 mA to 152.5 mA, 67.4 mA to 157.4 mA, respectively. For +3°, the threshold current is 90.7 mA, and the ending current point is 153.7 mA. For -3°, the beginning oscillation current point is 70.6 mA, and the ending current point is 149.6 mA. The laser oscillation behavior at ±3° differs slightly from that at other angles, due to the reduced external cavity feedback efficiency. This reduction is induced as the incident light increasingly deviates from the optical axis of the corner-cube retroreflector. There are only a few current points that can establish oscillation, and the driving current ranges that allow for laser oscillation are no longer continuous. Moreover, there is a significant difference in the oscillation thresholds for the two rotation angles, because the +3° and -3° angles may not be symmetrical to each other. If we change the working temperature of FADOF, the Faraday laser can have more current points that can attain oscillation at ±3°, and even larger rotation angle range for laser oscillation, which means that +3° or -3° is not the misalignment range limit. The Faraday laser can also be established through another cavity shape, which the corner-cube retroreflector is placed face to face at the same optical axis with laser diode, and we believe the rotation angle range that can attain laser oscillation will also be improved. All these results provide strong evidence of the high mechanical robustness of this Faraday laser.

As for the mode hopping, it will also occur when the laser is operated under different rotation angles. At 0° position, the maximum no mode hop range is about 783MHz, similar with the result shown in Fig. 3(a). While for operational rotation angels, such as +1°, -2°, +2° and +3°, the maximum mode hopping frequency are within the range form 1.4 GHz to 1.6 GHz (from 852.358 nm to 852.354 nm). Through the analysis of the transmission spectrum (F=4 region) shown in Fig. 2(b), the wavelength around 852.358 nm corresponds to the left peak area of the zero point in Fig. 2(b), and the wavelength around 852.354 nm corresponds to the right peak area of the zero point in Fig. 2(b). The free spectral range (FSR) of the laser in this condition is around 428 MHz, smaller than the approximately 652 MHz in Fig. 3(a), and there will be more laser modes to improve the probability of being closer to the highest transmittance point. As a result, the adjacent laser modes will not be the only choice that can used to establish laser oscillation. For the remaining operational rotation angles, the wavelength around 852.358 nm is not appeared, it maybe because of the different efficiency with different power inside the cavity, having a little influence on the transmission spectrum. From the Fig. 4(a), we can also see that even the mode hopping occurs for different operational rotation angles, the output frequency can still keep near the atomic transition lines. Compared to Faraday lasers that used plane mirrors previously [15,16,18,24,32], it is evident that the greatly improved mechanical robustness provided by the corner-cube enhances the environmental compatibility and reliability of our Faraday laser. Additionally, when combined with the immunity to variations in the driving current and working temperature of the laser diode, this leads to hope for the next generation of ECDLs. Figure 4(b) shows the wavelength results when the only variable is the rotation angle of the corner-cube retroreflector. We can see that laser output is consistently maintained even when the rotation angle changes.

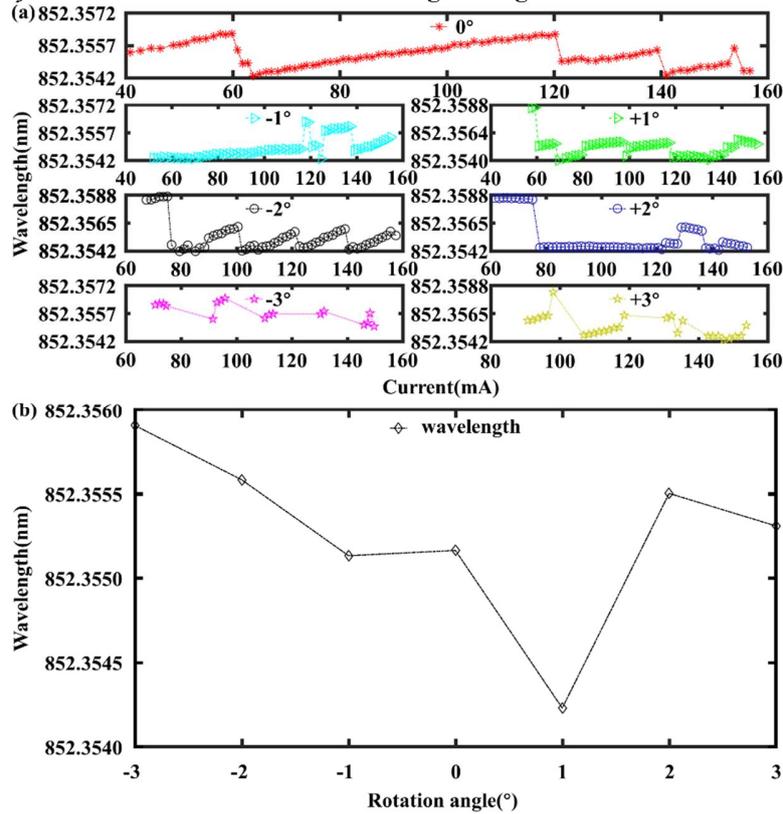

Fig. 4. (a) The current-wavelength curve for different rotation angles of Faraday laser. The rotation angle range is from -3° to +3°, and we can see that benefiting from the corner-cube retroreflector, the Faraday laser can still keep working even the detuning range is up to +3° or -3°. (b) The rotation angle-wavelength curve when the rotation angle of the corner-cube retroflector is the only variable.

Figure 5(a) illustrates the range of wavelength drift over time under different rotation angles. We maintained the working temperature and driving current of the ARLD constant to monitor the long-term frequency stability of the laser at various rotation angles. The working temperature of FADOF was also kept at the same for every angle to directly get the advantages of corner-cube retroreflector. When the working environment remains constant apart from the rotation angle, the Faraday laser can still function normally, highlighting a remarkable improvement of mechanical robustness and reliability. There is no mode hopping for at least 6 hours at -2° and the maximum wavelength range is ±1.2 pm at +1°. The output wavelength at different rotation angles closely aligns with the Cs $6S_{1/2}$, $F=4 \rightarrow 6P_{3/2}$, $F=5$ which is called the cycling transition, and we can easily get the SAS through the tunability of this Faraday laser.

Therefore, we can easily use the precision spectroscopy, such as modulation transfer spectroscopy [29,30], polarization spectroscopy [33,34], and dichroic atomic vapor laser lock [35,36] to stabilize the frequency of Faraday laser easily. In future works, we will do these to acquire a Faraday laser with high frequency stability and use it in applications such as atomic clocks, and atomic magnetometers for improvement of the system performance. The Pound-Drever-Hall system can also be used to suppress the frequency drift of Faraday lasers, we will combine it with the precision spectroscopy method to take advantage of the frequency stability properties at different time scales [37,38]. Fig. 5(b) displays the output mode of the Faraday laser in this study. Each point depicted in Figs. 3, 4(a) and 5(a) corresponds to laser oscillation with single-mode output, as illustrated in Fig. 5(b).

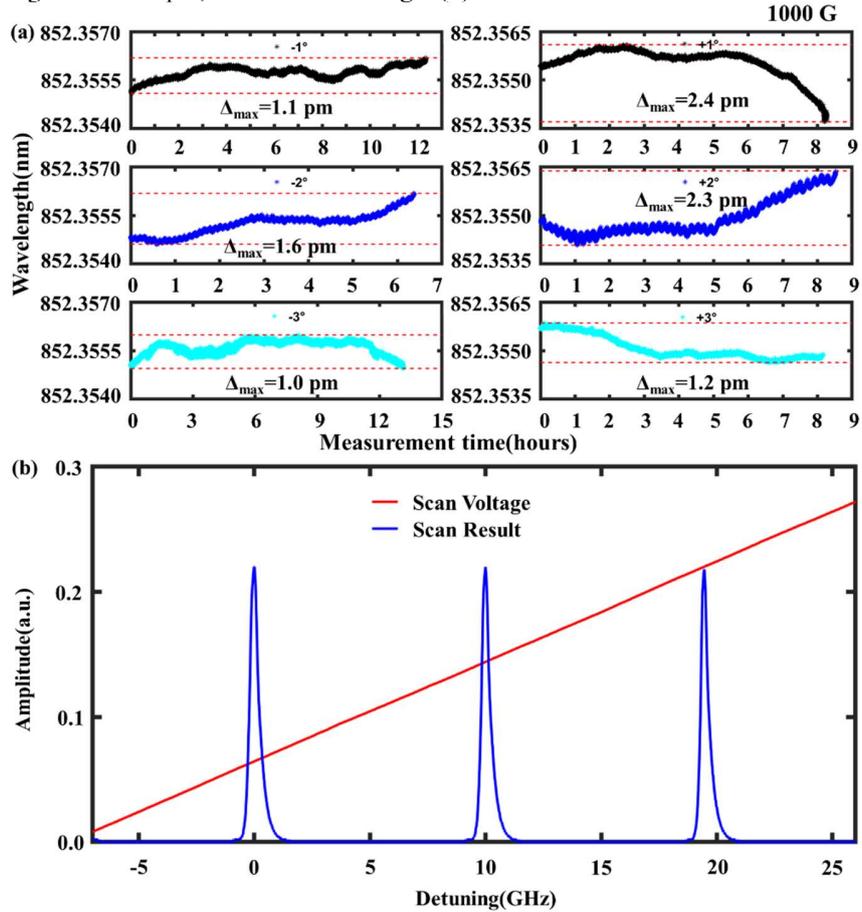

Fig. 5. (a) The wavelength range with time under different rotation angles. All the measurement results under different rotation angles are with the same driving currents and the same FADOF temperatures. We fixed the temperature of FADOF to 55.6°C, and also fixed the driving current of ARLD to 114.1 mA as well as working temperature of ARLD to 26.7°C. (b) The output mode scanned by a Fabry-Pérot interferometer, and each operation point in Fig. 3, Fig. 4(a) and Fig. 5(a) is single mode operation. At 0°angle in Fig. 4(a), the Faraday laser can achieve single-mode output within the widest current range from 40.8 mA to 156.6 mA.

## C. Linewidth and Allan deviation

We built another Faraday laser with the same parameters of the FADOF, and it also used a corner-cube retroreflector as the cavity mirror. The signal, generated through the process of heterodyne beating between the two Faraday lasers, is subsequently detected by a photodiode

(HAMAMATSU, C5658). The most probable linewidth and the frequency stability of the Faraday laser can be measured. The measurement results of beat signal by the spectrum analyzer (Keysight, N9010B) and the frequency counter (Keysight, 53230A) are shown in Fig. 6.

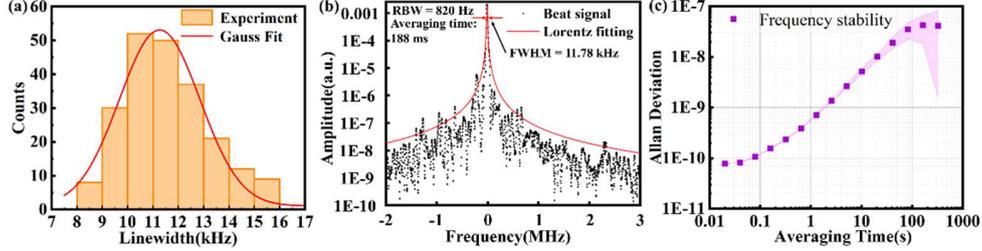

Fig. 6. Linewidth and frequency stability measured by heterodyne beating. (a) The histogram of repeated measurements of fitted beating linewidth between two home-made Faraday lasers whose working magnetic strength are both 1000 G. The other Faraday laser is also using the corner-cube retroreflector as the cavity mirror and the cavity length is about 15.6 cm. (b) The typical beating signal of two Faraday lasers. The Lorentz-fitted FWHM of the beat linewidth is 11.78 kHz. (c) The Allan deviation of the beating signal between the two Faraday lasers.

From Fig. 6(a), we can see that the most probable beating linewidth between two free running lasers is calculated to be 11.27 kHz. Assuming that these two lasers are independent and have the equal contribution to beating linewidth, the linewidth of each laser is 8 kHz, which is narrower than all the single-mode Faraday laser that has been reported to our best knowledge. Fig. 6(b) presents the typical heterodyne beating signal obtained from the two Faraday lasers, as recorded by the spectrum analyzer. The analyzer was set with a sweep time of 188 ms and a span bandwidth of 5 MHz. Fig. 6(c) is the Allan deviation of the beating signal between the two Faraday lasers, the short-term frequency stability is about 5.7E-10@1s. The stability can be further improved to at least the $10^{-13}$ level after implementing frequency locking through precision spectroscopy methods, such as modulation transfer spectroscopy [39].

### D. Noise performance

Figure 7 shows the noise performance of the Faraday laser in this paper. For the relative intensity noise (RIN), it will be influenced by the spontaneous emission, so we set the driving current to 114.1 mA, lower than 155 mA for more reliable results. We detected the laser with a photodiode (Thorlabs, APD430A/M), then we can obtain the RIN through combining the results of the phase noise analyzer (Rohde&Schwarz, FSWP) and the digital multimeter (Keysight, 34461A).

Figure 7(a) is the results of the RIN for different working temperatures of the ARLD, and the incident light intensity to the photodiode for 12.2°C, 26.7°C and 32.7 °C are about 18.3 μW, 18.9 μW and 18.7 μW, respectively. During this measurement process, the working temperature of the ARLD is the only variable, thus it can have an influence on the RIN. For all the working temperatures we set, the typical values of RIN at 10 kHz are lower than -140dBc/Hz, which is better than our previous work [24, 32].

Figure 7(b) is the result of the phase noise through heterodyne beating when the working temperature of the ARLD is 26.7°C, it is also measured by the phase noise analyzer (Rohde&Schwarz, FSWP). Compared with the heterodyne beating result of the commercial ECDLs working in the same room, it can be lower when the offset frequency is bigger than $10^3$ Hz. Fig. 7(c) is the result of integrating the phase noise. Using the 1/π-integral linewidth method shown in equation (1) can integrate the measured phase noise starting from the highest measured frequency offset, 100 MHz, up to the frequency offset where the integrated value reaches 1/π rad² [40-42], which in this work is 9.3 kHz. And the laser linewidth can reach to 6.6 kHz, closer to the 8.0 kHz result obtained before. At 9.3 kHz, this point represents the random walk of the laser phase over a 2π interval since the phase noise is a single sideband noise measurement [42].

$$\phi_{rms}^2 = \int_{\Delta \nu_{1/\pi}}^{\infty} S_\varphi(\nu)df = \frac{1}{\pi}\left[\text{rad}^2\right] \quad (1)$$

$S_\varphi(\nu)$ is the phase noise of the laser, $f$ is the Fourier frequency and $\Delta\nu_{1/\pi}$ is the 1/π-integral linewidth.

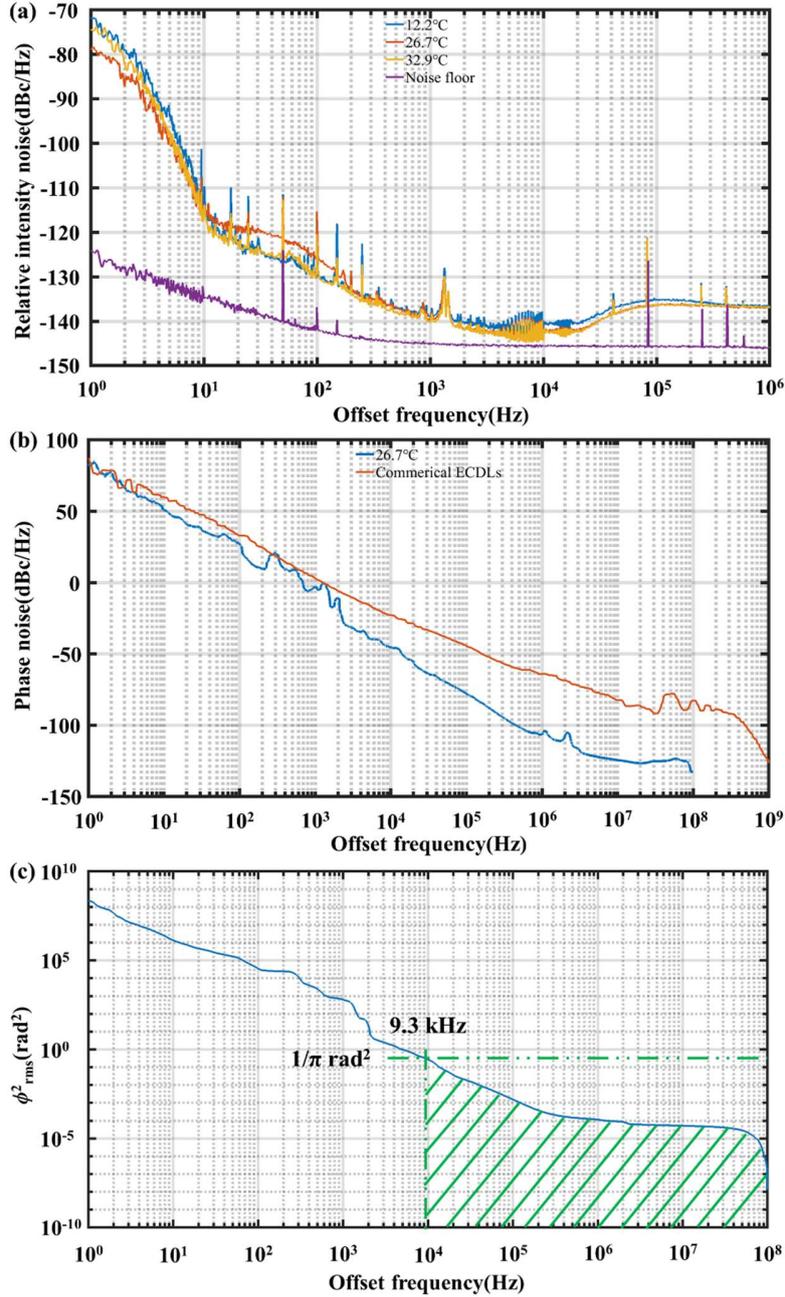

Fig. 7. Noise performance of the Faraday laser. (a) The relative intensity noise of the Faraday laser under different working temperature of the ARLD. (b) The phase noise of the free running Faraday laser through heterodyne beating of two lasers. (c) $1/\pi$-integral linewidth calculation for the Faraday laser.

To ensure the reproducibility of the corner-cube retroreflector as external cavity feedback element in Faraday laser for different working magnetic strength of FADOF, after the completion of 1000 G working magnetic strength experiment demonstrated in this paper, the FAODF will be dismantled and rebuilt to obtain another Faraday laser for other working magnetic field. We believe that it will have similar characteristics and performance.

## 5. Conclusions

In this work, we experimentally demonstrate a corner-cube-feedback Faraday laser. Benefiting from the robust mechanical structure, the laser oscillation can always be achieved unless the angle between incident light and the optical axis of corner cube is beyond the ±3° range. Through changing the working parameters such as temperature of atomic vapor cell in FADOF, the angle can exceed ±3°. Moreover, the output wavelength can directly correspond to the atomic transition lines. The wavelength fluctuation range is limited to ±1.2 pm within 9 hours under +3° rotation angles. Through heterodyne beating between two identical Faraday lasers, the most probable linewidth is 8 kHz, which is sufficiently narrow compared to most ECDLs. And for the further linewidth suppression to enhance the signal to noise of the optically pumped cesium beam resonator, this narrowed linewidth Faraday laser can be an alternative choice [43]. The narrower linewidth of the Faraday laser in this paper, compared to others, also allows for a reduced requirement in the electric locking bandwidth. The corner-cube retroreflector can also be used as feedback element in the excited-state Faraday laser [32], active Faraday optical frequency standard [44], high output power Faraday laser [26-28]. It is expected to improve their mechanical robustness and environmental compatibility in those applications. The corner-cube-feedback Faraday lasers, which with immunity to fluctuations of driving current and working temperature of the ARLD, combine with the higher mechanical robustness, have a wide application prospect in quantum precision measurement fields, including quantum optics, atomic physics, atomic magnetometers, cold atoms, and atomic gravimeters, etc.

## Funding

This work was funded by the China Postdoctoral Science Foundation (BX2021020), Wenzhou Major Science and Technology Innovation Key Project (ZG2020046), and Wenzhou Key Scientific and Technological Innovation R&D Project (2019ZG0029), Innovation Program for Quantum Science and Technology (2021ZD0303200).

## Disclosures

The authors declare no conflicts of interest.

## Data Availability Statement

The data that support the findings of this study are available from the corresponding author upon reasonable request.

## References

1. C. E. Wieman, L. Hollberg. "Using diode lasers for atomic physics," Rev. Sci. Instrum. **62**(1): 1-20 (1991).
2. E. C. Cook, P. J. Martin, T. L. Brown-Heft, J. C. Garman, and A. Steck. "High passive-stability diode-laser design for use in atomic-physics experiments," Rev. Sci. Instrum. **83**(4): 043101 (2012).
3. F. Riehle. "Frequency standards: basics and applications,". John Wiley & Sons (2006).
4. S. J. Seltzer. "Developments in alkali-metal atomic magnetometry," Princeton University, 2008.
5. D. Sesko, C. G. Fan, C. E. Wieman. "Production of a cold atomic vapor using diode-laser cooling," J. Opt. Soc. Am. B **5**(6): 1225-1227 (1988).
6. A. Peters, K. Y. Chung, S. Chu. "Measurement of gravitational acceleration by dropping atoms," Nature **400**(6747): 849-852 (1999).

7. R. Wyatt, W. J. Devlin. "10 kHz linewidth 1.5 μm InGaAsP external cavity laser with 55 nm tuning range," Electron. Lett. **19**(3): 110-112 (1983).
8. K. Liu, M. G. Littman. "Novel geometry for single-mode scanning of tunable lasers," Opt. Lett. **6**(3): 117-118 (1981).
9. S. Rauch, J. Sacher. "Compact Bragg grating stabilized ridge waveguide laser module with a power of 380 mW at 780 nm," IEEE Photonics Tech. L. **27**(16): 1737-1740 (2015).
10. B. E. Bernacki, P. R. Hemmer, S. P. Smith, and S. Ezekiel. "Alignment-insensitive technique for wideband tuning of an unmodified semiconductor laser," Opt. Lett. **13**(9): 725-727 (1988).
11. B. Dahmani, L. Hollberg, R. Drullinger. "Frequency stabilization of semiconductor lasers by resonant optical feedback," Opt. Lett. **12**(11): 876-878 (1987).
12. X. Baillard, A. Gauguet, S. Bize, P. Lemonde, P. Laurent, A. Clairon, and P. Rosenbusch. "Interference-filter-stabilized external-cavity diode lasers," Opt. Commun. **266**(2): 609-613 (2006).
13. M. Gilowski, C. Schubert, M. Zaiser, W. Herr, T. Wübbena, T. Wendirch, T. Müller, E. M. Rasel, and W. Ertmer. "Narrow bandwidth interference filter-stabilized diode laser systems for the manipulation of neutral atoms," Opt. Commun. **280**(2): 443-447 (2007).
14. Y. Ohman. "On some new auxiliary instruments in astrophysical research VI. A tentative monochromator for solar work based on the principle of selective magnetic rotation," Stockholms Obs. Ann, **19**(4): 9-11 (1956).
15. X. Miao, L. Yin, W. Zhuang, B. Luo, A. Dang, J. Chen, and H. Guo. "Note: Demonstration of an external-cavity diode laser system immune to current and temperature fluctuations," Rev. Sci. Instrum. **82**(8): 086106 (2011).
16. Z. Tao, Y. Hong, B. Luo, J. Chen, and H. Guo. "Diode laser operating on an atomic transition limited by an isotope 87 Rb Faraday filter at 780 nm," Opt. Lett. **40**(18): 4348-4351 (2015).
17. J. Keaveney, W. J. Hamlyn, C. S. Adams, and I. G. Hughes. "A single-mode external cavity diode laser using an intra-cavity atomic Faraday filter with short-term linewidth< 400 kHz and long-term stability of< 1 MHz," Rev. Sci. Instrum. **87**(9) (2016).
18. P. Chang, Y. Chen, H. Shang, X. Guan, H. Guo, J. Chen, and B. Luo. "A Faraday laser operating on Cs 852 nm transition". Applied Physics B, **125**(12): 230 (2019).
19. D. M. Camm, F. L. Curzon. "The Resonant Faraday Effect," Can. J. Phys. **50**(22): 2866-2880(1972).
20. X. Xue, D. Pan, X. Zhang, B. Luo, J Chen, and H. Guo. "Faraday anomalous dispersion optical filter at 133 Cs weak 459 nm transition," Photonics Res. **3**(5): 275-278 (2015).
21. M. A. Zentile, J. Keaveney, L. Weller, D. J. Whiting, and C. S. Adams. "ElecSus: A program to calculate the electric susceptibility of an atomic ensemble," Comput. Phys. Commun. **189**: 162-174 (2015).
22. J. Keaveney, C. S. Adams, I. G. Hughes. "ElecSus: Extension to arbitrary geometry magneto-optics," Comput. Phys. Commun. **224**: 311-324 (2018).
23. M. Ohtsu, Y. Teramachi, Y. Otsuka and A. Osaki. "Analyses of mode-hopping phenomena in an AlGaAs laser," IEEE J. Quantum Elect. **22**(4): 535-543 (1986).
24. T. Shi, X. Guan, P. Chang, J. Miao, D. Pan, B. Luo, H. Guo, and J. Chen. "A dual-frequency Faraday laser," IEEE Photonics J. **12**(4): 1-11, (2020).
25. J. Zhang, G. Gao, B. Wang, et al. "Background noise resistant underwater wireless optical communication using Faraday atomic line laser and filter," J. Lightwave Technol. **40**(1): 63-73 (2022).
26. M. Rotondaro, B. Zhdanov, M. Shaffer, and R. Knize. "Narrowband diode laser pump module for pumping alkali vapors," Opt. Express **26**(8): 9792-9797 (2018).
27. H. Tang, H. Zhao, R. Wang, L. Li, Z. Yang, H. Wang, W. Yang, K. Han, and X. Xu. "18W ultra-narrow diode laser absolutely locked to the Rb D 2 line," Opt. Express **29**(23): 38728-38736 (2021).
28. H. Tang, H. Zhao H, D. Zhang, L. Li, W. Yang, K. Han, Z. Yang, H. Wang, and X. Xu. "Polarization insensitive efficient ultra-narrow diode laser strictly locked by a Faraday filter," Opt. Express, **30**(16): 29772-29780 (2022).
29. P. Chang, H. Shi, J. Miao, T. Shi, D. Pan, B. Luo, H. Guo, and J. Chen. "Frequency-stabilized Faraday laser with $10^{-14}$ short-term instability for atomic clocks," Appl. Phys. Lett. **120**(14) (2022).
30. H. Shi, P. Chang, Z. Wang, Z. Liu, T. Shi, J. Chen. "Frequency Stabilization of a Cesium Faraday Laser With a Double-Layer Vapor Cell as Frequency Reference," IEEE Photonics J. **14**(6): 1-6 (2022).
31. E. R. Peck. "Theory of the Corner-Cube Interferometer," J. Opt. Soc. Am. **38**(12), 1015-1024 (1948).
32. P. Chang, H. Peng, S. Zhang, Z. Chen, B. Luo, J. Chen, and H. Guo. "A Faraday laser lasing on Rb 1529 nm transition", Scientific Reports **7**(1): 8995 (2017).
33. C. Wieman, T. W. Hänsch. "Doppler-free laser polarization spectroscopy," Phys. Rev. Lett. **36**(20): 1170 (1976).
34. M. L. Harris, C. S. Adams, S. L. Cornish, I. C. McLeod, E. Tarleton, and I. G. Hughes. "Polarization spectroscopy in rubidium and cesium," Phys. Rev. A **73**(6): 062509 (2006).
35. A. Millett-Sikking, I. G. Hughes, P. Tierney, and S. L. Cornish. "DAVLL lineshapes in atomic rubidium," J. Phys. B-At Mol. Opt. **40**(1): 187 (2006).
36. D. J. McCarron, I. G. Hughes, P. Tierney, and S. L. Cornish. "A heated vapor cell unit for dichroic atomic vapor laser lock in atomic rubidium," Rev. Sci. Instrum. **78**(9) (2007).
37. W. Zhang, L. Stern, D. Carlson, et al. "Ultranarrow linewidth photonic-atomic laser," Laser Photonics Rev. **14**(4): 1900293 (2020).

38. J. Sanjuan, K. Abich, L. Blümel L, et al. "Simultaneous laser frequency stabilization to an optical cavity and an iodine frequency reference," Opt. Lett. **46**(2): 360-363 (2021).
39. S. Lee, G. Moon, S. E. Park, et al. "Laser frequency stabilization in the $10^{-14}$ range via optimized modulation transfer spectroscopy on the 87 Rb D 2 line," Opt. Lett. **48**(4): 1020-1023 (2023).
40. D. R. Hjelme, A. R. Mickelson, R. G. Beausoleil. "Semiconductor laser stabilization by external optical feedback". IEEE J. Quantum Elect. **27**(3): 352-372 (1991).
41. W. Liang, V. S. Ilchenko, D. Eliyahu, A. A. Savchenkov, A. B. Matsko, D. Seidel, and L. Maleki. "Ultralow noise miniature external cavity semiconductor laser," Nat. Commun. **6**(1): 7371 (2015).
42. K. Liu, N. Chauhan, J. Wang, et al. "36 Hz integral linewidth laser based on a photonic integrated 4.0 m coil resonator," Optica, **9**(7): 770-775 (2022).
43. N. Dimarcq, V. Giordano, P. Cerez P, and G. Theobald. "Analysis of the noise sources in an optically pumped cesium beam resonator," IEEE T. Instrum. Meas. **42**(2): 115-120 (1993).
44. W. Zhuang, J. Chen. "Active Faraday optical frequency standard," Opt. Lett. **39**(21): 6339-6342 (2014).